\documentclass[prx,reprint,twocolumn,superscriptaddress,amsmath,amssymb,floatfix]{revtex4-1}

\usepackage[utf8]{inputenc}
\usepackage{graphicx}
\usepackage[usenames,dvipsnames]{xcolor}
\usepackage{epstopdf}
\usepackage{amsmath,amsthm,amssymb}
\usepackage{latexsym}
\usepackage{bm}
\usepackage{dcolumn}
\usepackage{braket}
\usepackage[normalem]{ulem}
\usepackage{times}

\newcommand{\ba}{\mbox{\boldmath$a$}}
\newcommand{\br}{\mbox{\boldmath$r$}}
\newcommand{\bR}{\mbox{\boldmath$R$}}
\newcommand{\bd}{\mbox{\boldmath$d$}}

\newcommand{\bE}{\mbox{\boldmath$E$}}
\newcommand{\bu}{\mbox{\boldmath$u$}}

\newcommand{\bG}{\mbox{\boldmath$G$}}
\newcommand{\bq}{\mbox{\boldmath$q$}}

\begin{document}
\bibliographystyle{apsrev4-1}

\title{Empty perovskites as Coulomb floppy networks: entropic elasticity and negative thermal expansion}

\author{Alexei V. Tkachenko}
\email{oleksiyt@bnl.gov}
\affiliation{CFN,
Brookhaven National Laboratory, Upton, NY 11973, USA}

\author{Igor A. Zaliznyak}
\email{zaliznyak@bnl.gov}
\affiliation{Condensed Matter Physics and Materials Science Division,
Brookhaven National Laboratory, Upton, NY 11973, USA}

\begin{abstract}
Floppy Networks (FNs) provide valuable insight into the origin of anomalous mechanical and thermal properties in soft matter systems, from polymers, rubber, and biomolecules to glasses and granular materials. Here, we use the very same FN concept to construct a \emph{quantitative} microscopic theory of empty perovskites, a family of crystals with ReO$_3$ structure, which exhibit a number of unusual properties. One remarkable example is ScF$_3$, which shows a near-zero-temperature structural instability and large negative thermal expansion (NTE).
We trace these effects to an FN-like crystalline architecture formed by strong nearest-neighbor bonds, which is stabilized by net electrostatic repulsion that plays a role similar to osmotic pressure in polymeric gels. NTE in these crystalline solids, which we conceptualize as Coulomb Floppy Networks, emerges from the tension effect of Coulomb repulsion combined with the FN's entropic elasticity, and has the same physical origin as in gels and rubber. Our theory provides an accurate, quantitative description of phonons, thermal expansion, compressibility, and structural phase diagram, all in excellent agreement with experiments. The entropic stabilization of critical soft modes, which play only a secondary role in NTE, explains the observed phase diagram. Significant entropic elasticity resolves the puzzle of a marked, $\approx$~50\% discrepancy between the experimentally observed bulk modulus and \emph{ab initio} calculations.
The Coulomb FN approach is potentially applicable to other important materials with markedly covalent bonds, from perovskite oxides to iron chalcogenides, whose anomalous vibrational and structural properties are still poorly understood.
\end{abstract}

{
}
\date{\today}

\maketitle

\section{Introduction}
Theoretical understanding  of condensed matter invariably rests on the concept of the hierarchy of energy scales associated with various intra- and inter-atomic interactions. Typically, the full microscopic Hamiltonian is reduced to an effective model, which describes the low-energy degrees of freedom, subject to the constraints imposed by the high-energy interactions. Floppy Networks (FNs) is a class of such  models, in which constituent particles are bound by rigid links and where the total number of constraints imposed by these links is smaller than  required for the global rigidity of the system, i.e. violating the Maxwell's criterion of mechanical stability. A simple and intuitive physics of under-constrained FNs has been widely used in  soft condensed matter, where it plays prominent role in explaining  anomalous mechanical and thermal properties of polymers, rubber, fibers, glasses, and granular materials \cite{Flory_Principles1953,Flory_book1969,deGennes_book1979,Tanaka_SciAm1981,AnnakaTanaka_Nature1991,Alexander_PhysRep1998,Wyart_PRL2008,Broedersz_NatPhys2011}. The defining feature of a FN is the existence of zero-energy deformations. These "floppy" modes lead to a number of unusual properties, including entropic elasticity, negative thermal expansion (NTE) \cite{Flory_book1969,deGennes_book1979}, fragile behavior \cite{Wyart_PRL2008,Mao_instability, Broedersz_NatPhys2011}, non-Gaussian fluctuations \cite{AT_TW}, and topological edge modes \cite{topoPNAS2012,topoPRL}.

In contrast, FN-type models are relatively rare \cite{SimonVarma_PRL2001,HeCvetkovicVarma_PRB2010} in the realm of traditional solid state physics. This is surprising, because interactions between the nearest atoms in many crystalline materials are significantly stronger and more rigid than all others, yet their number is often not sufficient for the global rigidity of the structure. This makes FN an excellent starting point for describing many complex crystalline solids. In this paper, we illustrate the potential of this approach by constructing a comprehensive microsopic theory of a family of open-framework ionic crystals, empty perovskites, which recently gained prominence thanks to the observation of large and tunable isotropic NTE, crucial for many technological applications \cite{Mary_Science1996,Sleight_AnnuRev1998,Ernst_Nature1998,Barrera_etal_JPCM2005,Lind_Materials2012,Dove_RepProgPhys2016,Morelock_JAP2013,Morelock_ChemMat2014,Morelock_JSolStChem2015,Hancock_ChemMater2015,Hu_MZrF6_JACS2016,Hester_ChemMat2017,Hester_InorgChem2018,Hester_JSolStChem2019}. One remarkable representative of this family is ScF$_3$ \cite{Greve_JACS2010,Handunkanda_PRB2015,Hu_JACS2016,Li_PRL2011,Wendt_2019}, where NTE is observed in the vicinity of a structural quantum phase transition: the cubic lattice becomes unstable under a very modest external pressure, often less than  1~GPa. In the absence of controlled microscopic theory, the question of the physical origin of NTE in ScF$_3$ remained controversial. The two leading contenders are quartic phonon anharmonicity \cite{Li_PRL2011} and the correlated anion vibrations known as rigid unit modes (RUMs) \cite{Dove_RepProgPhys2016}. In either case, ScF$_3$ with its simple, empty perovskite crystal structure [Fig.~\ref{Fig1:structure}(a)] has been proposed as a perfect example of the respective mechanism. Here, we develop a microscopic theory which shows that neither of these mechanisms is crucial for the observed anomalous behaviors, in agreement with the recent experiments \cite{Wendt_2019}.

In RUM models, the low-energy atomic vibrations are approximated by rotations, $\theta$, of rigid units formed by the coordination polyhedra [Fig.~\ref{Fig1:structure}(a)-(c)]. In ScF$_3$, a model structure of freely-jointed rigid ScF$_6$ octahedra is fully constrained \cite{Dove_RepProgPhys2016}. While this structure is unstable due to zero-energy RUM at the Brilluoin zone boundary [Fig.~\ref{Fig1:structure}(b)], the number of such floppy modes per unit cell vanishes in the thermodynamic limit, and so does macroscopic NTE. The effect, however, does appear in 1D  chain model [Fig.~\ref{Fig1:structure}(d)], which has two floppy modes per unit cell \cite{Dove_RepProgPhys2016} and in a variation of RUM model proposed in Refs. \cite{SimonVarma_PRL2001,HeCvetkovicVarma_PRB2010}, which in 2D has one floppy  mode per unit cell [Fig.~\ref{Fig1:structure}(c)].
In the absence of tension (negative pressure at the boundary), floppy modes make these RUM models unstable and the equilibrium, $\theta = \theta_0$, is phenomenologically enforced by postulating the bending rigidity of flexible joints with potential energy $U \sim (\theta - \theta_0)^2$.
Despite being only toy models, these  mechanistic models do capture qualitatively the rubber-like entropic nature of NTE and ``guitar string'' tension effect, as well as the role played by geometric constraints \cite{SimonVarma_PRL2001,HeCvetkovicVarma_PRB2010,Barron_AnnPhys1957,Barron_book1999}.

In our microscopic theory, ReO$_3$-type empty perovskites [Fig.~\ref{Fig1:structure}(e)] and other ionic compounds with FN architecture are described as Coulomb Floppy Networks (CFNs) with freely-jointed rigid links representing bonds between nearest neighbors.
The FN description directly connects to other well known examples of NTE materials, the polymeric gels and rubber \cite{Flory_book1969,Alexander_PhysRep1998,SimonVarma_PRL2001}. A CFN crystal [Fig ~\ref{Fig1:structure}(e)] can be viewed as a 3D cross-linked polymer structure. Just like gels are stabilized by osmotic pressure which applies tension to FN, the CFN solids are stabilized by internal negative pressure originating from ionic Coulomb repulsion [Fig.~\ref{Fig1:structure}(f),(g)]. 
Thanks to regular, periodic structure of crystalline solids, a quantitative  microscopic theory extending well beyond mechanistic toy models can be constructed around this concept. The theory provides an accurate quantitative description of fluctuational (entropic) elasticity, NTE, and the underlying structural phase transition, all in excellent agreement with the experimental observations \cite{Greve_JACS2010,Li_PRL2011,Handunkanda_PRB2015,Hu_JACS2016,Wendt_2019,Morelock_JAP2013,Morelock_ChemMat2014,Morelock_JSolStChem2015,Hancock_ChemMater2015,Hu_MZrF6_JACS2016,Hester_ChemMat2017,Hester_InorgChem2018,Hester_JSolStChem2019}.
While we focus at the specific case of cubic ScF$_3$, most of our conclusions can be adapted to other crystalline materials featuring FN structure. In fact, such is the situation in many systems of high topical interest, ranging from pure silicon, which shows NTE at low temperature to perovskite oxides and iron chalcogenides \cite{Kim_etal_Fultz_PNAS2018,Zhou_etal_Pindak_PNAS2010,Fobes_PRB2016} where strong, highly covalent nearest-neighbor bonds form under-constrained FNs.

\section{Coulomb Floppy Network Hamiltonian}
In constructing our theory, we capitalize on the clear separation of energy scales between those deformations of the lattice that do change the lengths of stiff nearest-neighbor Sc-F bonds and those that do not. We formalize this idea by representing all the interactions between the nearest-neighbor Sc and F ions with a single potential, $V_b(r_b)$. A convenient starting point is to assume that this potential is infinitely rigid, which corresponds to the FN limit where all Sc-F bonds have a fixed length, $r_b = r_0$. This assumption is later relaxed, leading to a controlled perturbation theory.

\begin{figure}
\includegraphics[width=.49\textwidth]{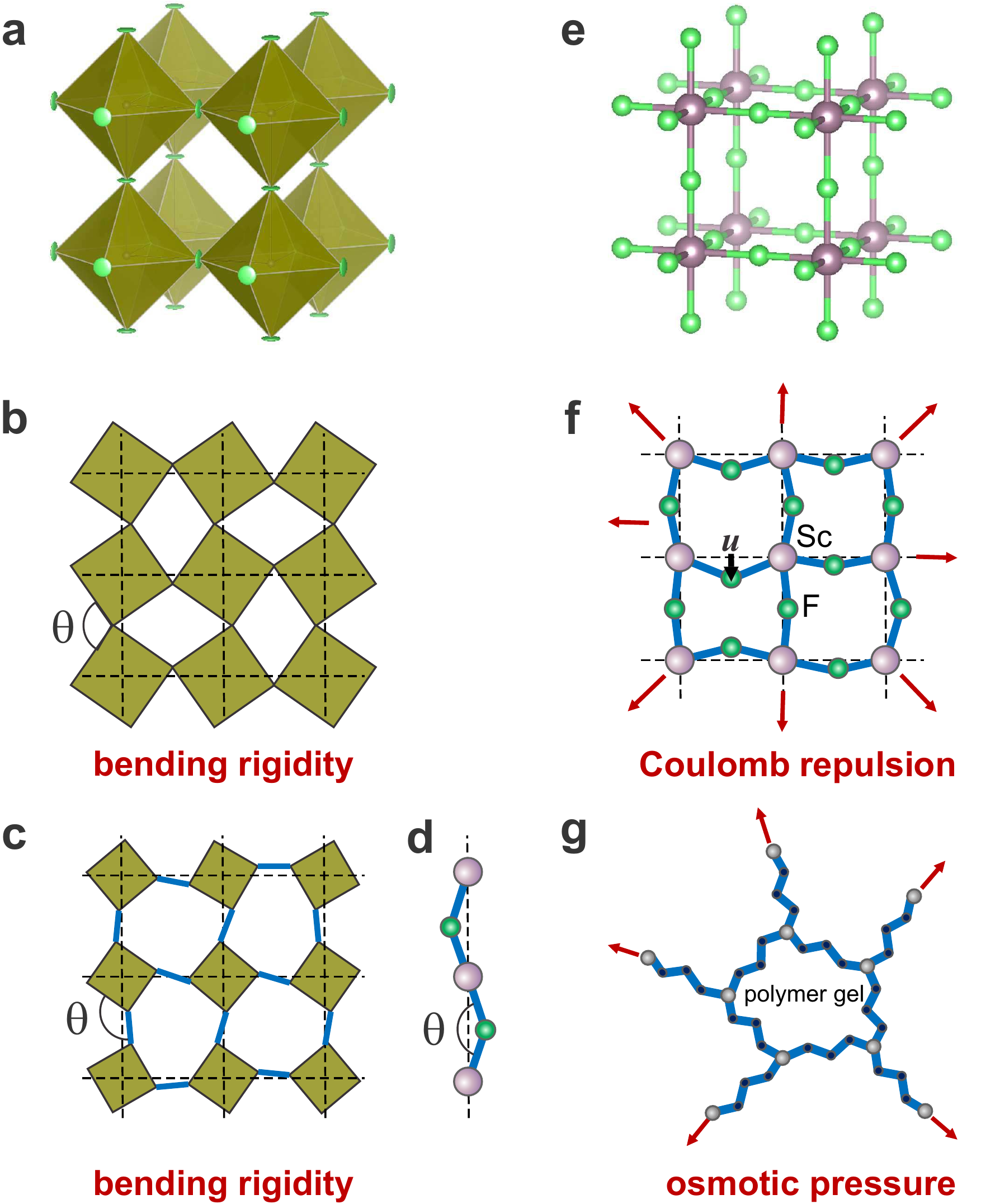}
\caption{{\bf Crystal structure of ScF$_3$ and schematics of RUM and Coulomb floppy network models.} (a) Coordination polyhedra rendering of the ReO$_3$ type cubic empty perovskite structure. (b) - (d) the RUM-based mechanistic models. Assuming rigid ScF$_6$ octahedra, or squares in the two-dimensional version (b), the structure is fully constrained, with no floppy modes. (c) A generalized model of Refs.~\onlinecite{SimonVarma_PRL2001,HeCvetkovicVarma_PRB2010} has one floppy mode, and (d) the one-dimensional chain model has two \cite{Dove_RepProgPhys2016}. In the absence of tension at equilibrium, $\theta = \theta_0$, the classical RUM models are unstable and are phenomenologically stabilized by postulating the bending rigidity of flexible joints with potential energy $U \sim (\theta - \theta_0)^2$. (f), (g) A freely-jointed floppy network structure in ScF$_3$ is stabilized by net Coulomb repulsion. At equilibrium, the network is under tension (negative pressure). Thermal motion of F ions (green spheres) transverse to rigid Sc-F bonds pulls Sc ions closer together and acts against Coulomb tension, leading to NTE. (h) In a polymer gel, a disordered floppy network is stabilized by internal osmotic pressure \cite{Flory_Principles1953,Flory_book1969,deGennes_book1979,Tanaka_SciAm1981}.
}
\label{Fig1:structure}
\end{figure}

In the absence of other interactions between the ions, our system would be a true floppy network: with 12 degrees of freedom and 6 constraints per unit cell, there would be 6 zero energy modes that do not change the lengths of Sc-F bonds. These "floppy" modes can be parameterized by displacements, ${\bu}_{n \nu}$, of F ions in the directions perpendicular to the corresponding  primitive vectors of the cubic lattice, ${\ba}_{\nu}$ ($n$ indexes the unit cells of the cubic lattice, $\nu= x,y,z$ indexes the position of the F ion in the unit cell, with Sc at the origin). In the leading order, there is no displacement of F along the Sc-F bond, i.e. ${\ba}_{\nu} \cdot {\bu}_{n \nu}=0$. If $r=a/2$ is  half of the lattice constant, it must  satisfy the following geometric constraint,
\begin{equation}
\label{rigid_bond}
r^2 -r_b^2+\langle{{\bu}_{n \nu}}^2\rangle = 0 \;.
\end{equation}
Here, $\langle{{\bu}_{n \nu}}^2\rangle$ is the thermal average, which in the thermodynamic limit is equivalent to the system average, $\langle{{\bu}_{n \nu}}^2\rangle=\frac{1}{3N}\sum_{n,\nu}{\bu}_{n \nu}^2$ ($3N$ is the number of F ions, $N \gg 1$).

The Coulomb energy of an ideal cubic ScF$_3$ lattice can be expressed as, $-3NM\tilde{e}^2/r$, where $M = 2.98$ is the Madelung constant, and effective static charges of Sc and F ions are $3\tilde{e}$ and $-\tilde{e}$, respectively. In an ideal ionic crystal, $\tilde{e} = e$ is electron charge; in real materials it is reduced by covalency. Since the electrostatic interactions between the nearest Sc and F ions are already included in $V_b$, we only need to account for the contributions from all other pairs of ions, which results in the electrostatic term, $3N(6-M)\tilde{e}^2/r$, in the Hamiltonian. In order to incorporate the geometric constraint, \eqref{rigid_bond}, into our theory, we use the standard method \cite{reif2009fundamentals}, introducing the auxiliary conjugated Lagrange multiplier field, $\kappa/2$, and adding the corresponding coupling term, $\delta H=\frac{\kappa}{2}\sum_{n,\nu}\left(r^2 -r_b^2+{\bu}_{n \nu}^2\right)$, to the effective Hamiltonian of the system,

\begin{equation}
\label{rigid_bond_Hamiltonian}
\begin{split}
{H}_{\perp} = {K} + 3N\left[\frac{(6-M)\tilde{e}^2}{r} + \frac{\kappa(r^2-r_b^2)}{2}\right] + 8Nr^3P \\
+ \frac{\kappa}{2}\sum_{n,\nu} {\bu}_{n \nu}^2 + \frac{\tilde{e}^2}{2r^3} \sum_{n,n',\nu,\nu'} {\bu}_{n \nu}\cdot \widehat{\bG}_{nn',\nu\nu'}\cdot {\bu}_{n' \nu'} \;.
\end{split}
\end{equation}
Here, ${K}$ is kinetic energy, $8Nr^3P = PV$ accounts for external pressure, and the last term describes the leading, second-order fluctuational correction to the electrostatic energy associated with the transverse displacements of F ions, ${\bu}_{n \nu}$.
Higher-order terms in ${\bu}_{n \nu}$ can also be straightforwardly obtained in the same spirit of multipole expansion. Interestingly, quartic term, which we considered, turns out to be \emph{negative} and thus destabilizing. However, it has a smallness of $\sim \langle {u}_{n \nu} \rangle^2 /r_0^2$, which is  $\sim 10^{-2}$ even at 1000~K \cite{Greve_JACS2010,Wendt_2019} and therefore can be safely neglected except, perhaps, in a close vicinity of the structural phase transition.

The electrostatic repulsion between non-nearest neighbor ions provides the tension force that stabilizes the structure and renders finite stiffness to the floppy modes. The equilibrium values of $\kappa$ and $r$ can be found by minimizing the free energy, or, equivalently, $\langle {H}_{eff}\rangle$. Minimization with respect to $\kappa$ simply recovers the original geometric constraint, Eq.~\eqref{rigid_bond}. On the other hand, minimization with respect to $r$ yields the Lagrange multiplier, $\kappa= (6-M)\tilde{e}^2/r^3 - 8rP $. This parameter determines the tensile force applied to each Sc-F-Sc link, $f = \kappa r/2$, which balances the negative internal pressure due to the net electrostatic repulsion. Its effect can be compared to the osmotic pressure in a polymer gel, which results in stretching of an otherwise floppy polymer network, making it rigid, as illustrated in Fig.~\ref{Fig1:structure}(f)-(g). The (positive) external pressure reduces the tension on Sc-F bonds and therefore shifts the system back towards the floppy network regime.

\section{Phonon Spectrum}
After switching in Eq.~\eqref{rigid_bond_Hamiltonian} to Fourier representation, ${\bu}_{\bf q \nu} = \frac{1}{\sqrt{N}} \sum_{n} {\bu}_{n \nu} e^{-i {\bq} \cdot \br_{n \nu} }$, and diagonalization of the dipole tensor, $\widehat{\bG}_{{\bf q}, \nu\nu'}$, the canonical Hamiltonian of uncoupled oscillators is obtained  (as shown in Appendix~\ref{point_charge}),
\begin{equation}
\label{H_diag}
{H}_{\perp} = \sum_{{\bf q},\sigma}\left[ \frac{|p^{(\sigma)}_{\bf q}|^2}{2m_F} +\frac{\tilde{e}^2}{2 r^3} \left(6-M - P/P_0 +\gamma^{(\sigma)}_{\bf q}\right)\left| {u}^{(\sigma)}_{\bf q}\right |^2 \right] .
\end{equation}
Here, $\gamma^{(\sigma)}_{\bf q}$ ($\sigma = 1,...,6$) are the eigenvalues of $\widehat{\bG}_{{\bf q}, \nu\nu'}$ obtained upon the diagonalization and ${u}^{(\sigma)}_{\bf q}$ are the respective eigenmodes; $P_0 = {\tilde{e}^2}/{8r_0^4} \approx 12.4$~GPa is the characteristic pressure (assuming the effective charge $\tilde{e}=0.84e$ estimated from  Pauling's electronegativity \cite{Wendt_2019} and experimental value of $r_0=2.01$). The energies of F transverse phonon modes obtained from this Hamiltonian are,
\begin{equation}
\label{dispersion}
\hbar \omega^{(\sigma)}_{\bf q} = \hbar \omega_0 \sqrt{6-M + \gamma^{(\sigma)}_{\bf q} - P/P_0} \;,
\end{equation}
where $\hbar \omega_0 = \hbar \tilde{e}/\sqrt{r^3m_F} \approx 16.5$~meV.

\begin{figure}[!t]
\includegraphics[width=.48\textwidth]{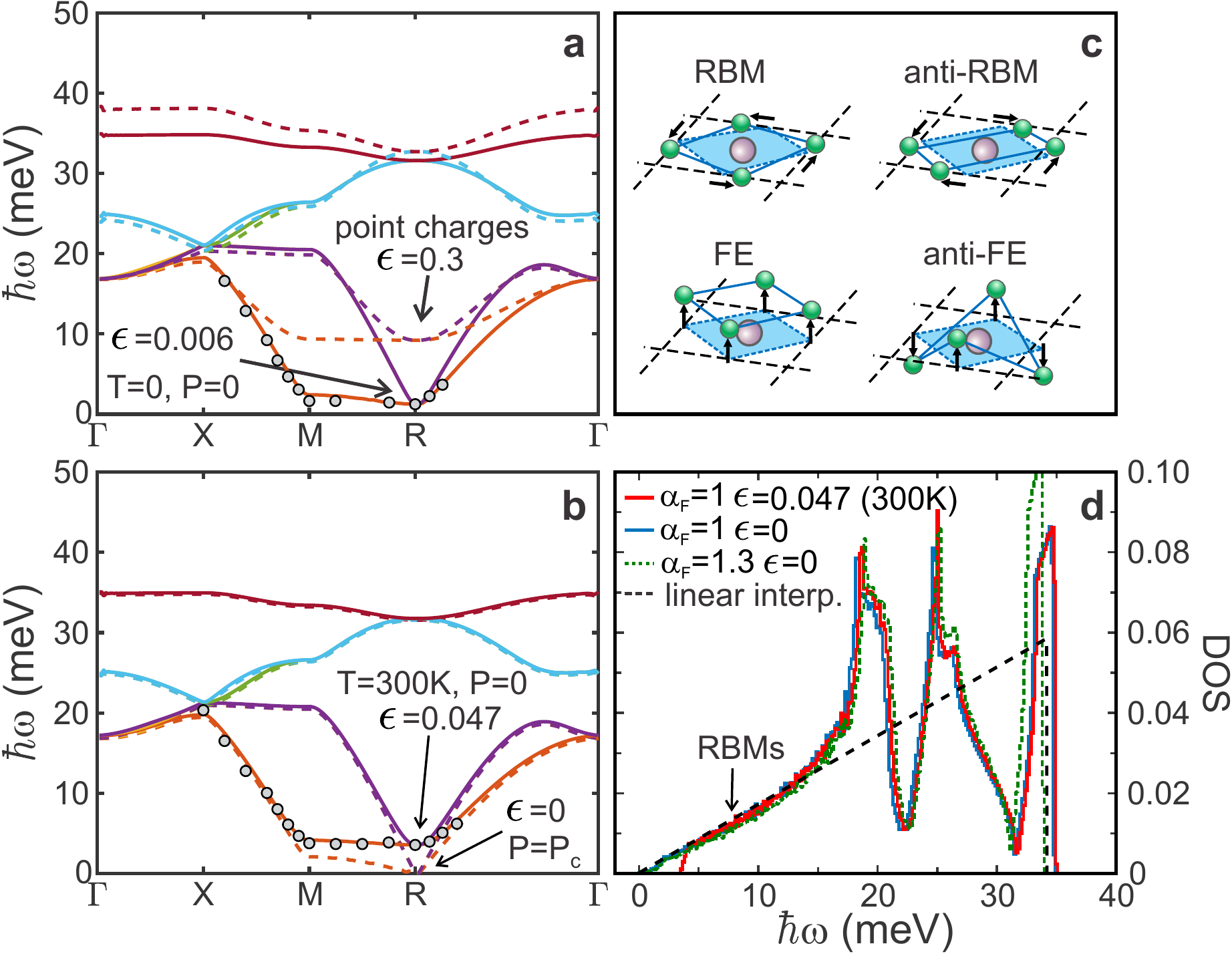}
\caption{{\bf Transverse phonons of Coulomb floppy network in ScF$_3$.} Curves are results of our theory, Eq.~\eqref{dispersion}, and symbols are the experimental data points for ScF$_3$ from Ref.~\cite{Handunkanda_PRB2015}. (a) The spectrum of the six transverse F phonon modes at $T = 0$, $P = 0$, calculated using point charge approximation ($\epsilon = 0.3$, dashed lines) and with the account for the ionic charge distribution ($\epsilon = 0.006$, solid lines). (b) The phonon spectrum at room temperature ($T = 300$~K, $P = 0$, $\epsilon = 0.047$, solid line), and at a critical point ($P = P_c(T)$, $\epsilon = 0$, dashed line). (c) Schematics of the F atom displacements for phonon eigenmodes described by Eqs.~\eqref{eigenmodes}: soft Rotating Bbreathing Mode (RBM) usually associated with quasi-RUMs, hard ferroelectric (FE) mode, and their counter-parts, anti-RBM and anti-FE modes. (d) The phonon density of states (DOS) for the six modes in (b), calculated for $\alpha_F = 1.0$ and $\alpha_F = 1.3$. Dashed line shows linear interpolation.
}
\label{Fig2:phonons}
\end{figure}

Adopting the point-charge approximation, i.e. neglecting electronic polarizability of the ions and the bond dipoles (which can be related to the difference between static, $\tilde{e}$, and dynamic, $ze$, (Born) effective charge \cite{BornChargeNote}, i.e. assuming $ze=\tilde{e}$), we obtain the dispersion curves shown by dashed lines in Fig.~\ref{Fig2:phonons}(a). The energy gap, $\hbar \omega_{\min} = \hbar \omega_0\sqrt{\epsilon_0-P/P_0}$, is controlled by parameter $\epsilon_0 = (6-M + \gamma_{\min})$, where $\gamma_{\min} = \min\{\gamma^{(\sigma)}_{\bf q}\}$. In a qualitative agreement with the experiment \cite{Greve_JACS2010,Handunkanda_PRB2015}, application of an external pressure reduces the soft mode gap, leading to a structural instability. Within the point-charge approximation, however, $\gamma_{\min} \approx -2.71$ and $\epsilon_0 \approx 0.31$, which results in a large gap, $\approx 10$~meV at $P = 0$, and critical pressure $P_c = \epsilon_0 P_0 \approx 4$~GPa, both of which are significantly higher than the experimental values.

Going beyond the point-charge approximation, one needs to account for electronic polarizabilities of F and Sc ions \cite{Wendt_2019}, $\alpha_F$ and $\alpha_{Sc}$, respectively, and for Sc-F bond polarizability, which is characterized by parameter $\delta < 1$. The parameter $\delta$ is related to a difference between the static and dynamic effective charges (see Appendix~\ref{ionic_dipole_correction} for details). These effects lead to the renormalization of $\gamma^{(\sigma)}_{\bf q}$ in Eq.~\eqref{H_diag},  $\gamma^{(\sigma)}_{\bf q} \rightarrow \gamma^{*(\sigma)}_{\bf q}$,
\begin{equation}
\label{gamma_q_renormalized}
\gamma^{*(\sigma)}_{\bf q} = \frac{\gamma^{(\sigma)}_{\bf q}\left[(1-\delta)^2 + \gamma_0\alpha_F/r^3\right] + \gamma_0\delta^2}{1 + (\gamma^{(\sigma)}_{\bf q} + \gamma_0) \alpha_F/r^3} ,
\end{equation}
and of the gap parameter, $\epsilon_0 = (6-M + \gamma^{*}_{\min})$, which depend on the effective charges and electronic polarizabilities.
The parameter $\delta$ is related to Born effective charge, $ze$, [Eq.~\eqref{Born_charge} in Appendix~\ref{ionic_dipole_correction}] and $\gamma_0 = -\sum_{\sigma} \gamma^{(\sigma)}_{\bf q}/6 \approx 0.90$ is obtained within the point-charge approximation \cite{Wendt_2019}. In principle, effective charges and electronic polarizabilities can be obtained from first principle calculations \cite{Curtarolo_NatMat2013,Oba_PhysRevMat2019}, or from experiment \cite{Shannon_PRB2006}.
However, in order to reliably determine parameter $\epsilon_0$ and thus the critical pressure, they need to be known with an exceptional precision. According to Eq.~\eqref{epsilon_0_renormalized}, if we assume the polarizability $\alpha_F = 1.0 \AA^3$, the spectral gap becomes zero at $ze/\tilde{e} \approx 0.994$, while a change to $\alpha_F = 1.3 \AA^3$ \cite{Wendt_2019,Shannon_PRB2006} could be offset by a $<2\%$ adjustment of the effective charge ratio, to $ze/\tilde{e} \approx 0.976$. This sensitivity provides an explanation for a wide variation of critical pressure and temperature among empty perovskite families chemically similar to ScF$_3$.

Using the experimental value of $P_c \approx 0.075$~GPa for ScF$_3$ at 0~K \cite{Greve_JACS2010}, we obtain $\epsilon_0 \approx 0.006$. Note, that determining this value with such an accuracy would require computation of the effective charges with a better than $0.5\%$ precision. Solid lines in Fig.~\ref{Fig2:phonons}(a) show the corresponding phonon dispersion curves. They are in excellent agreement with the experimentally measured dispersion of the low energy branch \cite{Handunkanda_PRB2015}, as well as the phonon spectra numerically computed in Ref.~\cite{Li_PRL2011}. The major difference is the absence of acoustic branches and hard optical phonons associated with longitudinal F oscillations, both of which are artificially "switched off"  in the rigid bond limit.

The most interesting feature of the obtained phonon spectra are the ultra-soft phonon modes at the edges of the Brillouin zone (BZ). These modes have a nearly flat dispersion along the M-R line, which reaches its global minimum at a vertex, ${\bq}^{*} = (\pm \pi/a, \pm \pi/a, \pm \pi/a)$, and are often interpreted as RUMs, a coordinated rotations of the rigid octahedra formed by Sc together with six surrounding F ions. The peculiar low-energy M-R dispersion of these modes, which determines the type of structural instability that occurs at ${\bq}^{*}$, originates from Coulomb interaction and is accurately described in our CFN theory. While this dispersion is properly captured by some of the DFT calculations \cite{Li_PRL2011}, it is often missed by DFT \cite{Curtarolo_NatMat2013,Oba_PhysRevMat2019}.
Although the ultra-soft floppy eigenmodes of the Hamiltonian~\eqref{rigid_bond_Hamiltonian} indeed resemble RUMs [but see Fig.~\ref{Fig2:phonons}(c) and discussion below], the fixed distance between the nearest F ions suggested by RUMs is not supported neither by numerical simulations \cite{Li_PRL2011}, nor by recent experiments \cite{Wendt_2019}. Furthermore, contrary to the common belief \cite{Dove_RepProgPhys2016} these putative ``RUMs'' are not markedly important for the thermomechanical properties, such as NTE. As we show below,
outside quantum regime (at $T \gtrsim 200$~K) these modes provide only sub-leading correction to the more na\"{i}ve Einstein model, which assumes completely un-coordinated vibrations of the neighbor atoms \cite{Wendt_2019}.

In fact, the geometry of the ultra-soft modes is fully determined by symmetry. For wave vectors at the edge ($q^{\mu}=q^{\nu}=\pi/a$), or at the center axis ($q^{\mu}=q^{\nu}=0$) of the Brillouin zone, which are invariant under symmetry transformations of the crystal that involve only $\mu$ and $\nu$ coordinates, the normal modes are either symmetric, or anti-symmetric with respect to the symmetry group of a square [($\nu, \mu, \tau$) is a triad of different coordinates]. This leads to the following general form for non-degenerate modes in those two cases,
\begin{equation}
\label{eigenmodes}
    u_{\pm}^{(\nu\mu)}({\bq})=
    \begin{cases}
    (u_{{\bf q},\tau}^{\nu}\pm u_{{\bf q},\tau}^{\mu}) / \sqrt{2}, &  q^\mu=q^\nu=0\\
    (u_{{\bf q},\mu}^{\nu}\pm u_{{\bf q},\nu}^{\mu}) /\sqrt{2}, & q^\mu=q^\nu=\pi/a
\end{cases} .
\end{equation}
Here, the minus sign corresponds to the lower energy branch. The ultra-soft modes on the $MR$ line are ``breathing'' rotations: the F ions are moving in a mid-plane between two Sc ions, somewhat resembling RUMs [Fig.~\ref{Fig2:phonons}(c)]. However, for a finite, even very large bond rigidity the Sc-F and F-F distances do change, $\sim \bu^2$. 
Despite being quadratic in $\bu$, these non-rigid deformations contribute to harmonic terms in the effective Hamiltonian, $H_\perp$: since Sc-F bond is under tension, its energy changes linearly with deformation, i.e. also $\sim \bu^2$. On the other hand, the finite Sc-F bond rigidity itself only contributes to the next order terms in $H_\perp$, giving rise to quartic anharmonicity \cite{Li_PRL2011}. These Rotating Breathing Modes (RBMs) provide the strongest reduction of the repulsive electrostatic energy between F ions, which offsets, at least partially, the energy penalty associated with tension of Sc-F bonds. Thanks to the cancelation of these two energy contributions, there is almost no restoring force associated with the F displacement for such a coordinated motion. Hence, RBMs are in fact \emph{the softest} of all the transverse modes of F ions. For $P = \epsilon_0 P_0$, the effects of Sc-F bond tension and F-F repulsion completely cancel each other, signalling an onset of a structural instability. Since the instability occurs at a vertex of the Brillouin zone, ${\bq}^{*}$, it involves simultaneous RBM displacements in all three planes and results in a transformation from cubic to rhombohedral crystal structure.

The four mode families given by Eq.~\eqref{eigenmodes} reveal the physics of the entire phonon spectrum. In the vicinity of ${\bq}=0$ ($\Gamma$ point), the six normal modes are represented by 3  ``ferroelectric'' and 3  ``anti-ferroelectric'' modes, with ion displacements along $x$, $y$, and $z$ direction, respectively. As one gradually moves away from the $\Gamma$ point, the phonons  can still be viewed as spatially modulated versions of those modes. The low-energy peak in the density of states, $\hbar \omega_1$, is  associated  with the soft "anti-ferroelectric" modes, while the higher-energy peaks, $\hbar \omega_2$ and $\hbar \omega_3$ correspond to "ferroelectric" modes (longitudinal and transverse, respectively).  In the vicinity of  $MR$ line, all modes are either ultra-soft RBMs or their high-energy counterparts,  anti-RBMs. In the density of states, the (spatially-modulated) RBMs and anti-RBMs appear as a linear shoulder below $\hbar \omega_1$ and above $\hbar \omega_2$, respectively.

\section{Entropic stabilization of criticality}
\label{stabilization_of_criticality}
Near the critical point, the floppy network behavior that has been suppressed by the electrostatic tension, is recovered. At a finite temperature, however, thermal fluctuations provide another, entropic mechanism for FN stabilization. Such entropic stabilization is aided by a steric constraint on displacement, $\bu_{n\nu}$, which was originally omitted in the effective Hamiltonian \eqref{rigid_bond_Hamiltonian}. Namely, the effect of core repulsion of the nearest F ions becomes significant when the fluctuations increase in amplitude on the way to become unstable according to the harmonic analysis. This can be accounted for by imposing a constraint, ${u_{n\nu}^{\mu}}^2 < u_0^2$, on each component, $\mu$, of the displacement field, ${\bu}_{n\nu}$, where $u_0 \sim 1$\AA\ is a model parameter which determines the effective steric confinement of F ions. The resulting partition function has the following form:
\begin {equation}
\begin{split}
Z={\bf Tr}\left( {e^{-\frac{{\hat H}_{eff}}{k_BT}} \prod_{n \nu \mu} {\Theta\left(u_0^2-{u_{n \nu}^{\mu 2}}\right)}}\right) = \\
={\bf Tr}\left(e^{-\frac{{\hat H}_{eff}}{k_BT}} \prod_{n \nu \mu} {\int\limits_{-i\infty}^{i\infty} \frac{e^{ \kappa_{n \nu}^{\mu} (u_0^2-{u_{n \nu}^{\mu 2}})}}{\kappa_{n \nu}^{\mu}} \frac{d\kappa_{n \nu}^{\mu}}{2\pi}}\right) .
\end{split}
\end{equation}

Here, we performed exponentiation of the step functions representing the steric constraints by introducing a fictitious Lagrange multiplier field, $\kappa_{n \nu}^{\mu}$ (formally, this is analogous to a common method of exponentiating a delta-function \cite{reif2009fundamentals}). Within the mean field approximation, all components of the lattice field $\kappa_{n \nu}^{\mu}$ can be replaced with a single value, $\kappa_T$. This results in the following addition to the effective Hamiltonian,
\begin{equation}
\label{H_steric}
\delta H_{F-F} = k_BT \sum_{n,\nu,\mu}\left[\kappa_T({u^{\mu 2}_{n \nu}}-u_0^2)+\ln \kappa_T \right] .
\end{equation}
By minimizing its thermal average with respect to $\kappa_T$, we obtain the equilibrium value of this parameter, $\kappa_T=1/(u_0^2-\langle {u^{\mu 2}_{n \nu}}\rangle) \approx 1/u_0^2$. We therefore conclude that the effect of a steric constraint can be represented as an additional rigidity, $\kappa_T=2k_BT/u_0^2$, adding $\delta H_T =  \sum_{n,\nu}\kappa_T{\bu}^{2}_{n \nu}/2$ to the quasi-harmonic Hamiltonian of the system, $H_\perp$. Combining the effects of temperature and pressure, we obtain the following expression for the gap parameter that controls the criticality,
\begin{equation}
\label{eps}
\epsilon(T,P) = \epsilon_0 +(\chi T -P)/P_0  \;,
\end{equation}
where $\chi = k_B/(4 r u_0^2)$. According to this result, there is a linear dependence of the critical temperature on pressure. It also predicts that the square of the floppy phonon's energy increases linearly with temperature, $(\hbar \omega(T))^2 = (\hbar \omega_0)^2 (\epsilon_0+\chi T/P_0)$, in perfect agreement with experiment \cite{Handunkanda_PRB2015}. By setting $u_0=1.00$~\AA, we obtain a close match between  the coefficient of proportionality, $(\hbar \omega_0)^2 \chi/P_0 = 2k_B\hbar^2/(m_Fu_0^2)\approx 0.038 $~meV$^2$/K, and its experimental value, $0.0376(5)$~meV$^2$/K \cite{Handunkanda_PRB2015}. This, in turn, allows us to evaluate the slope of the $P_c(T)$ curve, $dT_c/dP = \chi^{-1}\approx 580$~K/GPa, which compares favorably with the experimental value, $\approx 525$~K/GPa \cite{Greve_JACS2010,Handunkanda_PRB2015}. The resulting phase diagram, $P_c(T)$, is presented in the insert to Fig. \ref{Fig3:PhaseDiagram_NTE}(a). Fig. \ref{Fig3:PhaseDiagram_NTE}(a) shows the variation of critical temperature for Sc$_{1-x}$Ti$_x$F$_3$ and Sc$_{1-x}$Al$_x$F$_3$ series, which (aside from pure AlF$_3$) is quite accurately described by the $r$-dependence of $\epsilon_0$ given by Eq.~\eqref{epsilon_0_renormalized} in Appendix~\ref{ionic_dipole_correction}.

\section{NTE and entropic elasticity}
\label{NTE}
NTE emerges as a natural property of a floppy network: the constraint, \eqref{rigid_bond}, relates an increase in transverse fluctuations of F ions to the overall contraction of the crystal. 
As shown in Ref.~\cite{Wendt_2019}, the effect can already be evaluated within a simplified Einstein approximation, which completely ignores RBMs and the non-trivial dispersion of the transverse floppy modes. We can reproduce the result of Ref.~\cite{Wendt_2019} by replacing all eigenvalues $\gamma^{(\sigma)}_{\bf q}$ in \eqref{H_diag} with their average, $\gamma_0\approx 0.9$, obtained within the point charge approximation. In the classical regime, where according to the equipartition theorem the average potential energy of every mode equals $k_BT/2$, one obtains,
\begin{equation}
\label{alpha_E}
      \frac{r-r_b}{r_0} \approx-\frac{\langle {\bu}^2_{n \nu} \rangle}{2r_0^2}\approx -\frac{r_0k_BT}{{\Tilde e}^2(6-M - P/P_0 +\gamma_0)}=\alpha_ET.
\end{equation}
For $P=0$,  $\alpha_E=-\frac{r_0k_B}{{\Tilde e}^2(6-M +\gamma_0)}\approx -9 \cdot 10^{-6}K^{-1} $, which is surprisingly close to the experimental result for ScF$_3$ at room temperature. However, within the Einstein approximation the quantum effects would lead to freezing out of the floppy modes below $T\approx \hbar \omega_0 \sqrt{6-m+\gamma_0} /k_B \approx 270K$.

Here, we present a systematic calculation of the NTE effect with a full account for the dispersion of the transverse soft phonons. In order to calculate $\langle {\bu}^2_{n \nu} \rangle$, we first determine the density of states (DOS) from our theory. The DOS resulting from Eq.~\eqref{dispersion} features three peaks, at energies $\hbar \omega_1\approx 1.1\hbar \omega_0 \approx 18$~meV, $\hbar \omega_2\approx 1.5\hbar \omega_0 \approx 25$~meV, and $\hbar \omega_3\approx 2\hbar \omega_0 \approx 33$~meV [Fig. \ref{Fig2:phonons}(d)], in agreement with both numerical and experimental data \cite{Li_PRL2011}. Notably, the low energy RBM part of the DOS is well described by a linear function. This is because RBMs have very weak dispersion along the MR line. Their dispersion perpendicular to the MR line is thus approximately that of 2D phonons, which explains the linear DOS behavior.

We calculate NTE analytically by adopting a simple, approximate expression for the DOS, $g_\omega\approx \frac{2 \omega}{\lambda_+\omega_0^2} \Theta\left(\omega-\omega_-\right) \Theta\left(\omega_+-\omega\right)$, obtained by extrapolating the linear quasi-RUM behavior all the way up to the high frequency cut-off, $\omega_{+}=\omega_0\sqrt{\epsilon+\lambda_+}$. The opening of a gap for finite $\epsilon$ results in a low energy cut-off of the spectrum, at $\hbar\omega_-=\hbar\omega_0\sqrt{\epsilon+\lambda_-}$. Parameter $\lambda_-$ accounts for the effect of small, but finite dispersion of RBMs along the MR line, which makes the average energy along this line slightly higher than the gap, $\hbar\omega_0\sqrt{\epsilon}$. By performing the integration, $\langle {\bu}^2_{n \nu} \rangle=\hbar/2m_F\int d\omega  n_\omega g_\omega/\omega$, we obtain,
\begin{equation}
\label{alpha_Q}
     \Delta = \frac{r-r_b}{r_0} = -\frac{2r_0k_BT}{\tilde{e}^2\lambda_+}\ln \left(\frac{1-e^{-\hbar\omega_+/{k_BT}}}{1-e^{-\hbar \omega_-/{k_BT}}}\right) \,.
\end{equation}
In the classical regime, $k_B T \gtrsim \hbar \omega_{+}$, the logarithm in this expression can be replaced with $\ln(\omega_{+}/\omega_{-}) = \ln \sqrt{\lambda_{+}/(\epsilon + \lambda_{-})}$. The upper and lower cut-off parameters are obtained by matching this classical result with the one calculated for the exact DOS shown in Fig.~\ref{Fig2:phonons}(d). This gives  $\lambda_{+} \approx 4.3$ ($\hbar\omega_{+} \approx 35$~meV, $\hbar\omega_{+}/k_B\approx 400$~K), and $\lambda_{-}\approx 0.005$. As shown in Fig.~\ref{Fig2:phonons}(d), the resulting slope of $g_\omega$ is in excellent agreement with the linear portion of the exact DOS. The prefactor in front of the logarithm in Eq.~\eqref{alpha_Q} is very close to the result obtained within a much simper Einstein approximation, Eq.~\eqref{alpha_E} \cite{Wendt_2019}.
Hence, the non-trivial dispersion and ultra-soft RBMs are not essential for NTE and only give rise to a logarithmic correction factor ($\approx 2$ at $300$~K) to the Einstein model. The phonon dispersions gain importance in quantum regime, where Eq.~\eqref{alpha_Q} shows that the effect does not vanish below room temperature, contrary to the na\"{i}ve Einstein approximation. According to our result, NTE does not diverge even at the critical point, $\epsilon = 0$, when ultrasoft modes at the vertices R of the BZ become unstable. Such divergence (a putative argument for RUMs picture \cite{Dove_RepProgPhys2016}) is eliminated thanks to the non-vanishing Coulomb dispersion along the M-R line.

Another important property of floppy networks is entropic elasticity. A well-known example is elastic response of polymers and rubber to stretching, which suppresses the chain entropy \citep{Flory_Principles1953,Flory_book1969,deGennes_book1979}. In the case of a Coulomb FN solid, such as ScF$_3$, the overall compressibility remains finite even for infinitely rigid bonds thanks to thermal (and quantum) fluctuations. According to Eq.~\eqref{H_diag}, applied pressure reduces the stability of the network, increasing fluctuations and, therefore, entropy. The associated entropic contribution to bulk modulus can be calculated using Eqs.~\eqref{eps} and \eqref{alpha_Q}, $B_u=-\frac{1}{3}\left(\partial \Delta /\partial P \right)^{-1}=\frac{1}{3}\lambda_+P_0\tilde{e}^2(\epsilon+\lambda_-)/r_0k_BT$. Thanks to near-proportionality of $\epsilon$ to temperature, for finite T it quickly reaches a constant asymptotic value, $B_u\approx 2(r_0/u_0)^2\lambda_+P_0/3\approx 145$~GPa.

We can now relax the approximation of rigid bond with fixed length, $r_b$, by including the bond potential energy, which we expand up to the second order in deformation, $V_b(r_b)=V_b^{(0)}+ f_0(r_b-r_0)+\kappa_b (r_b-r_0)^2/2$, in the effective Hamiltonian \eqref{rigid_bond_Hamiltonian}. The geometric constraint, Eq.~\eqref{rigid_bond}, dictates that the two contributions to the compressibility, $B_b^{-1}=12r/(\kappa_b+\kappa)$ due to the finite bond rigidity and $B_u^{-1}$ associated with the fluctuations are additive, $B^{-1}=B_b^{-1}+B_u^{-1}$. This result explains large discrepancy between the DFT result for bulk modulus, $\approx 89$~GPa \cite{Curtarolo_NatMat2013,Oba_PhysRevMat2019}, and a much lower experimental value, $B  = 57(3)$~GPa, measured in ScF$_3$ at 300~K \cite{Greve_JACS2010,Morelock_JSolStChem2015}. Using the value $B_b = 89$~GPa from DFT and accounting for the fluctuational contribution to compressibility, we obtain $B=1/\left(B_b^{-1}+B_u^{-1}\right)\approx 55$~GPa, in excellent agreement both with experiment and molecular dynamics simulations \cite{Lazar_PRB2015}.

\begin{figure}[!t!h]
\includegraphics[width=.45\textwidth]{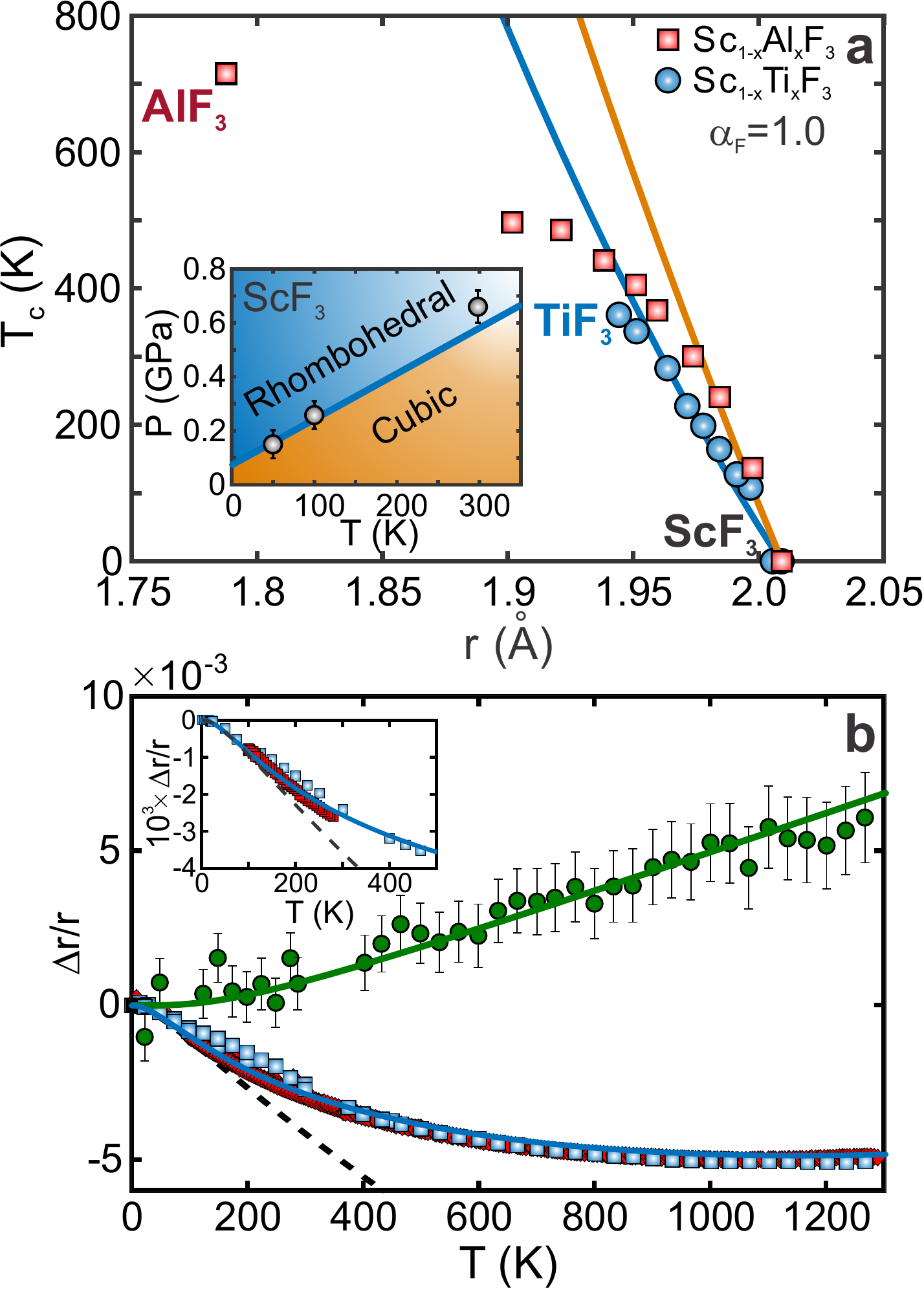}
\caption{{\bf Phase diagram and negative thermal expansion in ScF$_3$.} (a) Dependence of the critical temperature, $T_c$, on half of the lattice spacing, $r$. Symbols represent the experimental data for Sc$_{1-x}$Ti$_x$F$_3$ (circles) and Sc$_{1-x}$Al$_x$F$_3$ (squares) from Refs.~\cite{Morelock_ChemMat2014,Morelock_JSolStChem2015}. Solid lines are theoretical results for $\alpha_F = 1$ and Sc-F bond dipole parameter $\delta = \delta_0 + \delta_1 \Delta r/r_0$ (see Appendix~\ref{ionic_dipole_correction}) and include correction for the reduction of the confinement parameter with $r$, assuming $\Delta u_0 = \Delta r$. $\delta_0$ and $\delta_1$ were used as fitting parameters, resulting in $\delta_0 = 0.128$ and $\delta_1 = 0.35$ for Sc$_{1-x}$Ti$_x$F$_3$ and $\delta_1 = 0.20$ for Sc$_{1-x}$al$_x$F$_3$. The inset shows theoretical $P_c(T)$ phase diagram given by Eq.~\eqref{eps} with data points representing the experimental results for ScF$_3$ from Ref.~\cite{Greve_JACS2010}. (b) Solid lines show theoretical dependencies of the lattice NTE (blue), and Sc-F bond length (green). Symbols represent the experimental data from Refs.~\cite{Greve_JACS2010,Wendt_2019}. Dashed line is the result for infinitely rigid Sc-F bond, Eq.~\eqref{alpha_Q}.
}
\label{Fig3:PhaseDiagram_NTE}
\end{figure}

Our result for NTE, Eq.~\eqref{alpha_Q}, can be connected to the conventional theory of thermal expansion of solids, where,
\begin{equation}
\label{classics}
    \frac{r-r_0}{r_0} =\frac{1}{3VB}\sum_{i}\Gamma_{i}E_{i} .
\end{equation}
Here, $E_{i}$ are energies of individual phonon modes and $\Gamma_{i}$ are known as Gr\"{u}neisen parameters, which express the ratio between the contribution to internal pressure and the energy density for each mode,  $\Gamma_i=-\partial \ln{\omega_i} /\partial {\ln V}$. Typically, for crystalline solids Gr\"{u}neisen parameters are determined by the non-linearity of interatomic bond potentials and are of the order of 1. In the case of Coulomb FN in ScF$_3$, however, the dominant contribution, $\Gamma(\omega) \approx-B/2P_0(\omega_0/\omega)^2$, comes from the explicit pressure dependence of the Hamiltonian, $H_\perp$. Since $B \gg P_0$, this parameter is anomalously large in magnitude and negative. After substituting it into Eq.~\eqref{classics}, however, the bulk modulus cancels and the final result is indeed equivalent to Eq.~\eqref{alpha_Q}.

In our theory, we can employ Eq.~\eqref{classics} to describe the {\it positive} thermal expansion (PTE) of Sc-F bonds as an alternative to direct free energy minimization \cite{Wendt_2019}. In addition to the dominant contribution coming from the pressure dependence of phonon frequencies, there is a sub-dominant term $1/2$ in Gr\"{u}neisen parameters, $\Gamma_i$, resulting from the $\sim 1/r^3$ scaling of potential energy in Eq.~\eqref{H_diag}. One can show that it is this term that describes the phonon pressure responsible for PTE of Sc-F bond; the compressibility $B^{-1}$ in Eq.~\eqref{classics} in this case has to be replaced with $B_b^{-1}$ (see Appendix \ref{bond_expansion}). Thus, the very same floppy phonons that give rise to a pronounced NTE also lead to PTE of Sc-F bonds. As shown below, this approach also allows to phenomenologically account for the nonlinearity of Sc-F bond potential neglected so far.

We complete the description by recalling that there are 6 more phonon branches in the system, $3$ acoustic phonons and $3$ hard optical modes associated primarily with longitudinal displacements of Sc and F ions, which were neglected in the rigid bond picture. We account for these modes within Einstein approximation, by representing the deformation of 6 Sc-F bonds in response to the displacement of a Sc ion with a simple harmonic potential with spring constant $2V_b''+4V_b'/r_0 = 2(\kappa_b+\kappa) = 24Br_0$. The corresponding phonon energy is $\hbar \omega_{Sc}=\hbar\sqrt{24Br_0/m_{Sc}} \approx 50$~meV. The spring constant for longitudinal F oscillations, without account for electrostatic corrections, is $2\kappa_b = 2(12B-8P_0(6-M))r_0$, which gives the energy of hard optical phonon, $\hbar \omega_{\parallel} = \hbar\sqrt{24(B-2P_0)r_0/m_{F}}\approx 63$~meV. Both energies are very close to the peaks in phonon DOS observed experimentally, as well as to numerical results \cite{Li_PRL2011}. We account for the contribution of these modes to thermal expansion by assigning them a single energy, $\hbar \omega_{hard} = \hbar (\omega_{Sc}+\omega_{\parallel})/2\approx 56$~meV and an effective Gr\"{u}neisen parameter, $\Gamma$.

Combining the contributions from soft floppy transverse phonons and hard longitudinal modes we obtain,
\begin{equation}
\label{Delta_b}
 \Delta_b=\frac{r_b-r_0}{r_0}\approx\frac{k_BT} {B_bv_0}\left[D_2\left(\frac{\hbar\omega_{+}}{k_BT}\right)+2\Gamma \Phi\left(\frac{\hbar\omega_{hard}}{k_BT}\right) \right] .
\end{equation}
Here, $\Phi(x)=x/(\exp(x)-1)$ is the Bose-Einstein correction function to the equipartition theorem (Bernoulli function) and $D_2(x) = 2/x^2 \int_0^x x'^2(\exp(x')-1)^{-1} dx'$ is a 2D Debye function. The overall NTE effect is given by the sum of the earlier result obtained within rigid bond approximation, Eq.~\eqref{alpha_Q}, and Sc-F bond extension associated with internal phonon pressure given by Eq.~\eqref{classics} (see Appendix \ref{bond_expansion}),
\begin{equation}
\label{NTE_final}
\frac{r-r_0}{r_0} = \Delta + \frac{B_b}{B} \Delta_b .
\end{equation}
As shown in Fig.~\ref{Fig3:PhaseDiagram_NTE}(b), our results for NTE and for PTE of Sc-F bonds are in excellent agreement with experimental data in ScF$_3$. By using $\Gamma$ and the effective charge, $\tilde{e}$, as the only adjustable parameters, we obtain near-perfect fit of NTE for $\tilde{e}=0.81(3)e$, closely matching the Pauling estimate that we assumed earlier and the DFT results \cite{Bocharov_LTP2016}, and $\Gamma = 0.96(8)$, well within the expected range. The resulting curve for PTE of Sc-F bond also agrees very well with experiment [green solid line in Fig.~\ref{Fig3:PhaseDiagram_NTE}(b)].

\section{Conclusions}
In conclusion, we extended the concept of FN, which is ubiquitous in polymers, disordered and soft matter
\cite{Flory_Principles1953,Flory_book1969,deGennes_book1979,Tanaka_SciAm1981,AnnakaTanaka_Nature1991,Alexander_PhysRep1998,Wyart_PRL2008,Broedersz_NatPhys2011}, to open framework ionic solids. By doing so, we constructed a theory that treats these materials as Coulomb Floppy Networks stabilized by the net electrostatic repulsion and provides an accurate quantitative description of structural phase diagram and thermoelastic properties. We presented our approach by considering the specific example of ScF$_3$, which demonstrates the physics of CFNs unobscured by structural complexities \cite{Greve_JACS2010,Li_PRL2011,Handunkanda_PRB2015,Hu_JACS2016,Wendt_2019}. The tension in Sc-F-Sc bonds resulting from the net Coulomb repulsion leads to finite stiffness of transverse fluctuations and endows FN with structural stability. External pressure reduces the bond tension and, combined with the electrostatic energy gain due to coordinated RBM motion, leads to an instability of cubic structure. Thermal fluctuations under steric constraint, on the other hand, provide mechanism for entropic stabilization, leading to an increase of the spectral gap and the critical pressure with temperature.

NTE effect emerges as a natural consequence of the underlying floppy network behavior. Contrary to the common belief in the field, NTE does not rely neither on ultra-soft (quasi-)RUMs \cite{SimonVarma_PRL2001,HeCvetkovicVarma_PRB2010,Dove_RepProgPhys2016}, nor on quartic phonon anharmonicity, such as illustrated in \cite{Li_PRL2011}.
While softening of RBM phonons is indeed essential for the underlying structural instability and quartic corrections could be important for those ultra-soft modes in the vicinity of the critical point, a na\"{i}ve Einstein approximation, which ignores dispersion and softening of the floppy modes already gives a surprisingly good estimate of NTE and PTE of the Sc-F bond. The full account for the dispersion of the transverse floppy phonons presented here, including the ultra-soft RBMs, only yields a logarithmic correction to Einstein approximation. Still, that account is significant, as it explains why NTE does not vanish in the quantum regime below the room temperature where all the phonons within the simplified Einstein model are frozen out.

It is well known that thermal expansion of a solid is rooted in an anharmonicity of the Hamiltonian. In our case, the anharmonicity is geometric in origin and appears as a constraint \cite{SimonVarma_PRL2001,HeCvetkovicVarma_PRB2010}, leading to a theory that is essentially quasi-harmonic and starkly distinct from classical Gr\"{u}neisen-type theories where thermal expansion is driven by anharmonic inter-atomic potentials. Our effective quasi-harmonic Hamiltonian explicitly accounts for the anharmonicity of the electrostatic Coulomb interactions as well as nearest-neighbor steric (core) repulsion. The anharmonicity of the Sc-F bond potential, on the other hand, is included on a phenomenological level, within Gruneisen theory, as described above.

Our theory describes phonon spectra, phase diagram, NTE, and entropic elasticity of CFN crystals and is in excellent agreement with experimental data in ScF$_3$. It also explains the observed crystallographic phase diagrams of related crystal families with variable chemical composition [Fig.~\ref{Fig3:PhaseDiagram_NTE}(a)]. Beyond empty perovskites exemplified by ScF$_3$, our approach naturally extends to other perovskites, which is presently one of the most technologically important material families \cite{GuzmanVerri_Nature2019}. In order to describe these systems, one only needs to include an additional (``type-B'') ion at the center of the cubic unit cell. Our approach can also be applied to explain anomalous vibrational and structural properties of other topical materials, from pure silicon to cuprates and iron pnictides \cite{Kim_etal_Fultz_PNAS2018,Zhou_etal_Pindak_PNAS2010,Fobes_PRB2016}, and opens new avenues for predictive modeling of these effects in solids and metamaterials \cite{Nykypanchuk_Nature2008,Tkachenko_PRL2011}, which currently are beyond the reach of \emph{ab initio} methods.
Finally, the entropic correction to compressibility calculated in this work for the case of ScF$_3$ is likely to be substantial in a much wider range of materials. It is of marked importance from both conceptual and practical points of view, since the prediction of a bulk modulus is often used for validation of first-principle methods.


\begin{acknowledgments}
We gratefully acknowledge discussions with B. Fultz, A. Abanov, M. Hybertsen, P. Allen, and J. Hancock.
We also thank C. Varma for the interest to our work and useful comments. 
This work at Brookhaven National Laboratory was supported by Office of Basic Energy Sciences (BES), Division of Materials Sciences and Engineering,  U.S. Department of Energy (DOE),  under contract DE-SC0012704. Work at BNL's Center for Functional Nanomaterials (CFN) was sponsored by the Scientific User Facilities Division, Office of Basic Energy Sciences, U.S. Department of Energy, under the same contract. \\
\end{acknowledgments}


\begin{thebibliography}{52}
\bibitem [{Bor()}]{BornChargeNote}%
  \BibitemOpen
  \href@noop {} { {\bibinfo {title} {The dynamic effective charge,
  $ze$, which relates ionic displacement with the resulting dipole moment (also
  known as Born effective charge) is in a general case different from the
  static charge, $\tilde{e}$, that enters the Madelung energy}}\ }\BibitemShut
  {NoStop}%
\end{thebibliography}


%


\pagebreak

\begin{widetext}
\section*{Appendices}
\begin{appendix}

\section{Electrostatic coupling within point charge approximation}
\label{point_charge}

Within the point charge approximation, the electrostatic contribution to the Hamiltonian, ${H}_\perp$, in terms of "floppy" mode F displacements, ${\bu}_{n \nu}$, can be written as follows,
\begin{equation}
\label{Vel_SI}
V_{el} = \frac{\Tilde{e}^2 } {2r^3} \sum_{n,n',\nu,\nu'}{\bu}_{n \nu}\cdot {\bf\widehat{G}}_{nn',\nu\nu'}\cdot {\bu}_{n'  \nu'} = \frac{\Tilde{e}^2 } {2r^3} \sum_{n, \nu} \left[-\gamma_0 \left| {\bu}_{n \nu} \right|^2 + \sum_{(n', \nu')\neq (n,\nu)}{\bu}_{n \nu}\cdot{\bf{\widehat T}}_{nn',\nu\nu'}\cdot {\bu}_{n' \nu'}\right] .
\end{equation}
Here, $a = 2r$ is the lattice constant, $n,n'$ index the lattice unit cells, summations are over the F positions in the lattice (Sc displacements are neglected, $\nu, \nu' = x,y,z$), and
\begin{equation}
{T}^{\mu\mu'}_{nn',\nu\nu'} = -r^3\frac{\partial^2 R^{-1}}{\partial R^\mu \partial R^{\mu'}} =r^3\left(\frac{\delta^{\mu\mu'}}{R^3} - \frac{3 {R}^\mu {R}^{\mu'}}{R^5}\right) , {\bR}={\br}_{n\nu}-{\br}_{n'\nu'} ,
\end{equation}
is the traceless dipole tensor ($\sum_\mu {T}^{\mu\mu}_{nn',\nu\nu'} = 0$) where $\mu, \mu' = x,y,z$. On account of the F site symmetry, the diagonal force constant for transverse displacements, $\gamma_0$, is given by ($\mu \neq \nu$),
\begin{equation}
\gamma_0 = \sum_{n', {\nu'}(non-NN)} z_{\nu'} { T}^{\mu\mu}_{nn',\nu\nu'}  = -\frac{1}{2}\sum_{n', {\nu'}(non-NN)} z_{\nu'} { T}^{\nu\nu}_{nn',\nu\nu'} \approx 0.90 , \nu' = 0,x,y,z .
\end{equation}
Note, that here the summation over $\nu'$ includes the terms coming from interaction with Sc ions (to which $\nu' = 0$ is assigned), with the exception of the nearest Sc neighbors of a given F. Accordingly, $z_{\nu'} = +3$ for $\nu'=0$, and $z_{\nu'} = -1$ for $\nu' = x,y,z$.

After switching to the Fourier variables defined in the main text, ${\bu}_{\bf q \nu} = \frac{1}{\sqrt{N}} \sum_{n} {\bu}_{n \nu} e^{-i {\bq} \cdot \br_{n \nu} }$, the non-local part of the electrostatic potential can be re-written in the following form,
\begin{equation}
\sum_{(n',\nu')\neq (n,\nu)}{\bu}_{n \nu}\cdot{\bf{\widehat T}}_{nn',\nu\nu'}\cdot {\bu}_{n' \nu'}= \sum_{{\bf q}, \nu, \nu'}  {\bu}_{{\bf q}\nu }\cdot {\bf{\widehat T}}_{{\bf q},\nu \nu'}\cdot {\bu}_{{\bf -q}\nu'}
\end{equation}
\begin{equation}
{T}_{{\bf q},\nu \nu'}^{\mu\mu'}=\frac{3a^3}{8}\left(\frac{{\partial }^2}{{\partial q^\mu}{\partial q^{\mu'}}}-\frac{\delta^{\mu\mu'}}{3}\frac{{\partial }^2}{{\partial \bf{q}}\cdot {\partial \bf{q}}}\right)\sum_{{\bR}_{\nu \nu'}} \frac {\exp(i {\bq}\cdot{\bR}_{\nu\nu'})}{R_{\nu\nu'}^5} \, .
\end{equation}
Here, ${\bR}_{\nu\nu'}  =({\ba}_\nu - {\ba}_{\nu'})/2 + k{\ba}_x + l{\ba}_y + m{\ba}_z$, and $k,l,m\in {\rm Z}$.

In order to simplify the diagonalization of tensor ${\bf{\widehat T}}_{{\bf q}, \nu\nu'}$, we first regroup the displacements, ${\bu}_{{\bf q}, \nu}$, into three 2D vectors, ${\bu}'_{{\bf q}, \nu} = \left( u_{{\bf q}, \nu}^\mu, u_{{\bf q}, \nu}^{\mu'} \right)$. Here, the triad of indices, $(\nu,\mu,\mu')$, is an even ($\epsilon_{\nu\mu\mu'}=1$) permutation of $(x,y,z)$. In this representation, each  matrix element, ${\bf{\widehat T}}_{{\bf q}, \nu\nu'}$, is a $2 \times 2 $ matrix,
\begin{eqnarray}
{\bf \widehat T}_{{\bf q}, \nu\nu}=
\frac{3}{8}\sum_{k,l,m \in {\rm Z}}\left( \begin{array}{c c}
 \frac{L^2}{3}-m^2  & \frac{{\partial }^2}{{\partial \phi^\mu}{\partial \phi^{\mu'}}}\\
\frac{{\partial }^2}{{\partial \phi^\mu}{\partial \phi^{\mu'}}} & \frac{L^2}{3}-l^2 \\
\end{array}
\right)\frac{\cos(k\phi^\nu)\cos(m\phi^\mu)\cos(l\phi^{\mu'})}{L^5} , \\
{\bf \widehat T}_{{\bf q}, \nu\mu}=\frac{3}{8}\sum_{k,l,m\in {\rm Z}} \left(\begin{array}{c c}
\frac{{\partial }^2}{{\partial \phi^\mu}{\partial \phi^{\mu'}}} & \frac{{\partial }^2}{{\partial \phi^\nu}{\partial \phi^\mu}} \\
 \frac{{L'}^2}{3}-l^2 & \frac{{\partial }^2}{{\partial \phi^\nu}{\partial \phi^{\mu'}}} \end{array}\right)
\frac{\cos((k+1/2)\phi^\nu)\cos((m+1/2)\phi^\mu)\cos(l\phi^{\mu'})}{{L'}^5} ,
\end{eqnarray}
where ${\phi}^\mu=a{q}^\mu$, $L=\sqrt{k^2+m^2+l^2}$, and ${L'}=\sqrt{(k+1/2)^2+(m+1/2)^2+l^2}$.

Diagonalization of ${\bf{\widehat T}}_{{\bf q}, \nu \nu'}$ for a given wave vector, ${\bq}$, yields six normal modes with the corresponding eigenvalues, $\Tilde{\gamma}_{\bf q}^{(\sigma)} = \gamma_{\bf q}^{(\sigma)} + \gamma_0$.

\section{Corrections to the point-charge approximation}
\label{ionic_dipole_correction}

In order to go beyond point-charge approximation, one needs to consider the actual distribution of the ionic charge. This distribution can be represented by a multipole expansion about the effective point charge position, which is identified with the position of ionic nucleus. The monopole terms in this expansion yield point charge approximation considered above. Here, we consider the contribution to the electrostatic energy of the next, dipole terms. The importance of dipole interactions for the electrostatics and soft mode in perovskites had been pointed out long ago by P.~W.~Anderson \cite{Anderson_1960}. In general, dipole terms result from two additional effects discussed in the main text: finite electronic polarizabilities of ions,  $\alpha_F$ and $\alpha_{Sc}$, and the difference between the static and dynamic effective charges, which accounts for nearest-neighbor bond polarization.

Within point charge approximation, displacement of a single F ion, ${\bu}_{n\nu}$, results in a local dipole moment, ${\bd}_{n\nu}^{(0)} = -\tilde{e}{\bu}_{n\nu}$. In a real material system, however, this result has to be corrected for non-trivial distribution and redistribution of ionic charges \cite{WarhurstWhittle_Nature1951,Mulliken_JChemPhys1962,Bader_JChemPhys1987}. This correction is included in the definition of Born effective charge (dynamic charge), $ze$, which determines the dipole moment resulting from a displacement, ${\bd}_{n\nu} = -ze{\bu}_{n\nu}$. In principle, $ze$ can be determined numerically, e.g. by DFT calculations \cite{Mulliken_JChemPhys1962}. In the spirit of our theory, which separates effects of the nearest-neighbor and the longer-range Coulomb interactions, we will distinguish between the corresponding two contributions to the dipole moment, ${\bd}_{n\nu} = {\bd}_{n\nu}^{(0)} + \tilde{e} \delta {\bu}_{n\nu} + {\bd}^*_{n\nu}$. Here, ${\bd}^*_{n\nu}$ is an induced dipole due to a local electric field from non-nearest neighbors and $\tilde{e} \delta {\bu}_{n\nu}$ is a dipole resulting from the polarization of the nearest-neighbor, Sc-F bond. A na\"{i}ve ionic model where this dipole moment is given by F polarization in the electric field of the nearest Sc ions gives $\delta = 6 \alpha_F/{r^3}$. This result, however, is an over-estimate as covalency effects that govern charge distribution in Sc-F bond will reduce the polarization na\"{i}vely expected for an ionic model \cite{WarhurstWhittle_Nature1951,Mulliken_JChemPhys1962,Bader_JChemPhys1987}. We therefore treat $\delta$ as a material parameter, which we expand to first order in bond length, $\delta = \delta_0 + \delta_1(r_b-r_0)/r_0$. As shown below, $\delta$ can be related to Born effective charge and thus can be obtained from \emph{ab initio} calculations, or determined experimentally. 

According to \eqref{Vel_SI}, for a specific case when only one F ion is displaced the point-charge electric field of other ions at the displaced position is, ${\bE}_{n\nu} = - \gamma_0 \tilde{e}{\bu}_{n\nu}/r^3$. Hence, ${\bd}^*_{n\nu} = - \alpha_F \gamma_0 \tilde{e}{\bu}_{n\nu}/r^3$ and Born effective charge, in a crude approximation that neglects displacements of all other ions and treats them as point charges is given by ${z}e = \tilde{e}(1-\delta + \alpha_F\gamma_0/r^3)$.
In a general case, however, multiple ions are displaced. The individual induced dipoles, ${\bd}^*_{n\nu}$, are then determined by minimization of a Hamiltonian, which includes the point-charge electrostatic energy of Eq.~\eqref{Vel_SI} and the energy associated with the electronic dipole polarization,
\begin{equation}
\label{H_d}
H_{d} = \sum_{n,\nu} \left(  {\bd}^{*2}_{n\nu}/{2\alpha_F} - \left(\tilde{e}\delta {\bu}_{n\nu} + {\bd}^{*}_{n\nu}\right){\bE}_{n\nu} \right) .
\end{equation}
Here,
\begin{equation}
\label{E_d}
\bE_{n\nu} = - \gamma_0 \frac{\tilde{e}}{r^3} \bu_{n\nu} + \frac{1}{2r^3} \sum_{n',\nu'} {\bf{\widehat T}}_{nn',\nu\nu'} \left( (1-\delta) \tilde{e} \bu_{n'\nu'} - {\bd}^{*}_{n'\nu'} \right) ,
\end{equation}
is local electric field at the displaced position, including the field of point charges and the additional dipole contribution. Without the latter, the corresponding electric field potential at the displaced position yields the point-charge energy \eqref{Vel_SI}. 
We thus have,
\begin{equation}
\label{Vel_d}
V_{el}^{d} = \frac{\tilde{e}}{2r^3} \sum_{n,\nu} \left[-\gamma_0 \tilde{e} {\bu}_{n\nu}^2 + \sum_{(n',\nu')\neq (n,\nu)}{\bu}_{n\nu}\cdot{\bf{\widehat T}}_{nn',\nu\nu'}\cdot \left( (1-\delta)\tilde{e}{\bu}_{n'\nu'} - {\bd}^{*}_{n'\nu'} \right) \right] + H_{d} .
\end{equation}
%
%
In terms of ${\bd}_{n\nu}^{(0)} = -\tilde{e}\bu_{n\nu}$ and total induced dipole moment, ${\bd}_{n\nu} = \tilde{e}\delta {\bu}_{n\nu} + {\bd}^{*}_{n\nu}$, we obtain,
\begin{equation}
\label{Vel_dipole}
\begin{split}
V_{el}^{d} = \frac{1}{2r^3} & \sum_{n,\nu} \left( \gamma_0 {\bd}_{n\nu}^2 + \frac{r^3}{\alpha_F} \left( {\bd}_{n\nu}^{(0)} \delta + {\bd}_{n\nu} \right)^2 \right) + \\
& \frac{1} {2r^3} \sum_{(n', \nu')\neq (n,\nu)} \left( {\bd}_{n\nu}^{(0)} + {\bd}_{n\nu} \right) \cdot{\bf{\widehat G}}_{nn',\nu\nu'}\cdot \left( {\bd}_{n'\nu'}^{(0)} + {\bd}_{n'\nu'} \right)  = \sum_{{\bf q}, \sigma} h^{d}_{{\bf q},\sigma} .
\end{split}
\end{equation}
Here, $h^{d}_{{\bf q},\sigma}$ is the electrostatic part of the Hamiltonian, ${H}_\perp$, which replaces the point-charge expression, $h_{{\bf q},\sigma} = ({2r^3})^{-1} {\gamma}^{(\sigma)}_{\bf q} \left|\tilde{e} {\bu}^{(\sigma)}_{\bf q}\right|^2$, in Eq.~\eqref{H_diag}, upon account for ionic dipole moments,
%
%
%
\begin{equation}
\label{hq}
h^{d}_{{\bf q},\sigma} = \frac{\gamma_0}{2r^3} \left| {\bd}_{{\bf q}, \sigma} \right|^2 + \frac{1}{2\alpha_F}\left| {\bd}^{(0)}_{{\bf q}, \sigma} \delta + {\bd}_{{\bf q}, \sigma}\right|^2 + \frac{{\gamma}^{(\sigma)}_{\bf q}}{2r^3} \left|{\bd}^{(0)}_{{\bf q}, \sigma} + {\bd}_{{\bf q}, \sigma}\right|^2 .
\end{equation}
In this expression, ${\bd}^{(0)}_{{\bf q}, \sigma} = -\tilde{e}{\bu}^{(\sigma)}_{\bf q}$ and ${\bd}_{{\bf q}, \sigma}$ are related to ${\bd}_{n\nu}^{(0)}$ and ${\bd}_{n\nu}$ through the same linear transformation as ${\bu}^{(\sigma)}_{\bf q}$ to ${\bu}_{n\nu}$ (Appendix~\ref{point_charge}). By minimizing this energy with respect to ${\bd}_{{\bf q}, \sigma}$, we obtain,
%
\begin{equation}
\label{d_prop_u}
{\bd}_{{\bf q}, \sigma} = \frac{ {\gamma}^{(\sigma)}_{\bf q} + \delta \frac{r^3}{\alpha_F} }{ \gamma_0 +{\gamma}^{(\sigma)}_{\bf q} + \frac{r^3}{\alpha_F} } \tilde{e} {\bu}^{(\sigma)}_{\bf q} .
\end{equation}
%
%
%
Consequently, $\gamma^{(\sigma)}_{\bf q }$ in Eq.~\eqref{H_diag} of the main text has to be renormalized as follows (Eq.~\eqref{gamma_q_renormalized} in main text),
%
%
\begin{equation}
\label{gamma_q_renormalized_B}
\gamma^{*(\sigma)}_{\bf q } = \frac{\gamma^{(\sigma)}_{\bf q}\left[(1-\delta)^2 + \gamma_0\alpha_F/r^3\right] + \gamma_0\delta^2}{1 + (\gamma^{(\sigma)}_{\bf q} + \gamma_0) \alpha_F/r^3} .
\end{equation}
In particular, this correction leads to a significant change in the soft mode spectral gap compared to the result of point charge approximation,
\begin{equation}
\label{epsilon_0_renormalized}
\epsilon_0 = 6-M+\gamma_{\min} = 6 - M +  \frac{\gamma^{(0)}_{\min} \left( (1-\delta)^2 + \gamma_0 \alpha_F/r^3\right) + \gamma_0\delta^2}{1 + ({\gamma}^{(0)}_{\min} + \gamma_0) \alpha_F/r^3} .
\end{equation}

Finally, from minimization of Eq.~\eqref{hq} one can relate parameter $\delta$ to Born effective charge, $ze$,
\begin{equation}
\label{Born_charge}
\begin{split}
\frac{ze}{\Tilde{e}}=\sum_\sigma \frac{\partial ({\bd}^{(0)}_{\sigma, \bf 0}+{\bd}_{\sigma, \bf 0})}{\partial {\bd}^{(0)}_{n\nu}}=1-\frac{1}{6}\sum_{\sigma=1..6}\frac{\delta+\alpha_F\gamma^{(\sigma)}_{\bf 0}/r^3}{1+\alpha_F(\gamma^{(\sigma)}_{\bf 0}+\gamma_0)/r^3}=\\
=1+\frac{(\gamma_0-\pi)\alpha_F/r^3-\delta}{6(1+\pi\alpha_F/r^3)}+\frac{(\pi/3+\gamma_0)\alpha_F/r^3-\delta}{2(1-(\pi/3)\alpha_F/r^3)}+\frac{\gamma_0\alpha_F/r^3-\delta}{3} ,
\end{split}
\end{equation}
where we used an analytic result for ${\bf q}=0$, $\sigma = 1,...,6$ eigenvalues: $(\gamma^{(\sigma)}_{\bf 0}+\gamma_0)=(\pi,0,0,\pi/3,\pi/3,\pi/3)$.

\section{Thermal expansion of ScF bond and fluctuational correction to NTE result}
\label{bond_expansion}

Upon relaxing the approximation of infinitely rigid Sc-F bonds and expanding the bond potential, $V_b$, in terms of $r_b-r_0$, the effective Hamiltonian takes the following form,
\begin{equation}
\label{flexible_bond_Hamiltonian}
H_{eff} =  3N\left[\frac{(6-M)\tilde{e}^2}{r} + \frac{\kappa(r^2-r_b^2)}{2}\right] +f_0(r_b-r_0)+ \frac{\kappa_b}{2} (r_b-r_0)^2 + H_\perp+H_{\parallel} .
\end{equation}
Here, $f_0 = \frac{(6-M)\tilde{e}^2}{2r_0^2}$ and $H_\parallel$ accounts for acoustic modes and optical phonons associated with displacements of F ions along Sc-F-Sc bonds. The fact that Sc-F bonds are under tension resulting from Coulomb repulsion has important consequences. As before, the tension in Sc-F bond is balancing the negative electrostatic pressure, but it is now also related to the extension of the bond, $f=f_0+\kappa_b(r_b-r_0)$. At non-zero temperature and pressure, we once again perform minimization of $\langle {H}_{eff} \rangle$, this time with respect to $r_b$. In the leading order in $\langle {\bu}^2_{n \nu} \rangle$ this gives,
\begin{equation}
\label{r_b}
\Delta_b\equiv \frac{r_b-r_0}{r_0} = \frac{4r_0}{\kappa_b +\kappa} \left(\frac{1}{8r_0^3}\sum_{{\bf q},\sigma} \left\langle \frac{\tilde{e}^2}{2r_0^3} \left(6-M + \gamma^{(\sigma)}_{\bf q}\right) \left| {u}^{(\sigma)}_{\bf q} \right|^2 \right\rangle -  P \right) =\frac{\Tilde{P}-P}{3B_b}\;,
\end{equation}
Here, $B_b=(\kappa_b +\kappa)/12r_0$ is the contribution to the bulk modulus associated with bond rigidity and $\Tilde{P}=\frac{1}{2}\langle \hat{H}_\perp \rangle /V$ has a meaning of phonon contribution to (negative) internal pressure. This result is an analogue of the classical Gr\"{u}neisen formula, but for Sc-F bond expansion. The prefactor $1/2$ has a meaning of effective Gr\"{u}neisen parameter for our transverse F modes, coming from $1/r^3$ scaling of their energies. As discussed in the main text, this is only a subdominant contribution to the total Gr\"{u}neisen coefficient that determines the overall NTE effect,
\begin{equation}
\label{Gamma_q}
\Gamma_{\bf q}^{(\sigma)}=\frac{1}{2}\left(1-\frac{V}{\epsilon + \gamma^{(\sigma)}_{\bf q}-\gamma_{\min}}\frac{\partial P}{\partial V}\frac{\partial \epsilon}{\partial P}\right) = \frac{1}{2}-\frac{B}{2P_0} \frac{\omega_0^2}{\omega^{(\sigma) 2}_{\bf q}} .
\end{equation}

Within our theory, in order to find the overall thermal expansion, $(r-r_0)/r_0$, we need to add the results for $\Delta$ and $\Delta_b$ given by Eqs.~(\ref{alpha_Q}) and (\ref{Delta_b}) of the main text. However, the value of $\Delta$, which depends on pressure via $\epsilon$, has to be adjusted due to internal phonon pressure, $\Tilde{P}$,
\begin{equation}
\label{final_NTE}
\frac{r-r_0}{r_0}=\Delta-\frac{\partial \Delta}{\partial P}\Tilde{P}+\Delta_b=\Delta+\left(\frac{1}{B_u}+\frac{1}{B_b} \right)\frac{\Tilde{P}}{3}= \Delta+\frac{B_b}{B}\Delta_b
\end{equation}
This result is now fully consistent with the traditional Gr\"{u}neisen formula, Eq.~\eqref{classics} of the main text, with $\Gamma_{\bf q}^{(\sigma)}$ given by Eq.~\eqref{Gamma_q}. The effect of all other phonons can be included in Eqs.~(\ref{r_b}), (\ref{final_NTE}) by adding the respective contributions to $\tilde{P}$. Including the two longitudinal Sc-F bond-stretching optic phonons, yields Eq.~\eqref{Delta_b} and the theoretical curves shown in Fig.~\ref{Fig3:PhaseDiagram_NTE}(b) of the main text.

We note, that the contributions of F transverse phonons to thermal expansion of Sc-F bond and NTE [Eqs.~(\ref{r_b}), (\ref{Gamma_q})] are obtained in our theory in a quasi-harmonic approximation. However, our effective quasi-harmonic Hamiltonian explicitly accounts for the anharmonicity of the electrostatic Coulomb interactions as well as nearest-neighbor steric (core) repulsion, which in fact underlie non-zero thermal expansion. The anharmonicity of the Sc-F bond potential, on the other hand, is included in Eq.~\eqref{Delta_b} on a phenomenological level within Gruneisen theory, as described above. These modes are neglected in the rigid-bond approximation, which holds below about 400~K where the bond-stretching phonons are frozen out and their anharmonicity is irrelevant. The effective Gruneisen parameter, $\Gamma \approx 1$, obtained for these phonons by fitting thermal expansion [Fig.~\ref{Fig3:PhaseDiagram_NTE}(b) of the main text] shows that this anharmonicity is weak.

\end{appendix}
\end{widetext}\

\end{document}